\documentclass[aps,pre,twocolumn,groupedaddress,showkeys]{revtex4-2}
\usepackage{graphicx,caption}
\usepackage{dcolumn}
\usepackage{bm}
\usepackage{hyperref}

\begin{document}

\title{C-A test of DNA force fields}


\author{Ivan A. Strelnikov}
\author{Natalya A. Kovaleva}
\author{Artem P. Klinov}
\author{Elena A. Zubova}
\email{zubova@chph.ras.ru}
\affiliation
{N.N. Semenov Federal Research Center for Chemical Physics, Russian Academy of Sciences, 4 Kosygin Street, Moscow 119991, Russia}
\date{\today}

\begin{abstract}
\noindent
The DNA duplex may be locally strongly bent in complexes with proteins, for example, with polymerases or in a nucleosome. At such bends, the DNA helix is locally in the non-canonical forms A (with a narrow major groove and a large amount of north sugars) or C (with a narrow minor groove and a large share of BII phosphates). To model the formation of such complexes by molecular dynamics methods, the force field is required to reproduce these conformational transitions for a naked DNA. We analyzed the available experimental data on the B-C and B-A transitions under the conditions easily implemented in modeling: in an aqueous NaCl solution. We selected six DNA duplexes which conformations at different salt concentrations are known reliably enough. At low salt concentrations, poly(GC) and poly(A) are in the B-form, classical and slightly shifted to the A-form, respectively. The duplexes ATAT and GGTATACC have a strong and salt concentration dependent bias toward the A-form. The polymers poly(AC) and poly(G) take the C- and A-forms, respectively, at high salt concentrations. The reproduction of the behavior of these oligomers can serve as a test for the balance of interactions between the base stacking and the conformational flexibility of the sugar-phosphate backbone in a DNA force field. We tested the AMBER bsc1 and CHARMM36 force fields and their hybrids, and we failed to reproduce the experiment. In all the force fields, the salt concentration dependence is very weak. The known B-philicity of the AMBER force field proved to result from the B-philicity of its excessively strong base stacking. In the CHARMM force field, the B-form is a result of a fragile balance between the A-philic base stacking (especially for G:C pairs) and the C-philic backbone. Finally, we analyzed some recent simulations of the LacI-, SOX-4-, and Sac7d-DNA complex formation in the framework of the AMBER force field.
\end{abstract}
\maketitle
\begin{figure}[h]
\begin{center}
\includegraphics[width=0.95\linewidth]{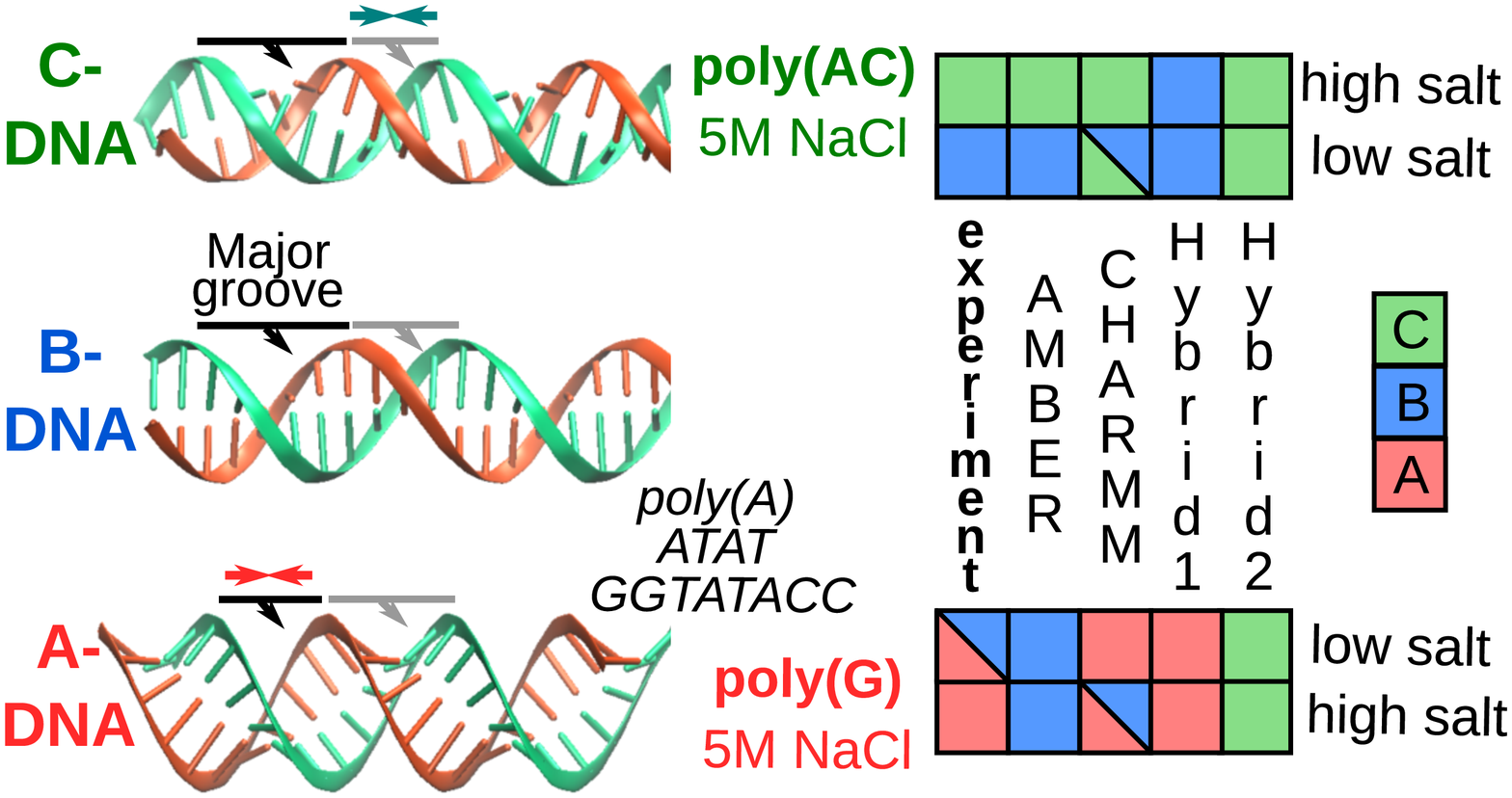}
\end{center}
\end{figure}
\section{Introduction}
\label{section-Intro}
\noindent
Understanding the mechanisms of the protein binding to DNA is a critical task not only in theoretical biophysics, but also in medicine. To date, there appeared a lot of reports on modeling the processes of DNA-protein binding or dissociation, protein diffusion and conformational dynamics in complexes (see reviews \cite{2015-DNA-protein-MD-binding-bending-folding-VanDerVaart,2020-MD-DNA-protein-review}). Usually, the researchers choose the AMBER force field. This force field allows simulating the complexes even in the case of highly deformed DNA (with kinks as a result of protein intercalation), for example, in complexes with Sox protein \cite{2011-DNA-protein-MD-trigger-DNA-conf-switch,2017-DNA-protein-binding-MD-Sox2-binding,2021-DNA-protein-binding-MD-rotation}, Hbb \cite{Hognon2019,Yoo2021}, Sac7d \cite{Spiriti2013,Zacharias2019,Singh2022},LacI \cite{2019-DNA-protein-binding-MD-A-DNA-lactose-repressor} and the others \cite{2014-DNA-protein-MD-TATAplus,2015-DNA-protein-binding-MD-kink-through-intercalation}. The CHARMM force field is being used much less frequently, with the important exception of nucleosomes \cite{Shaytan2016,Elbahnsi2018,Sun2022}. 

However, the reliability of the results of such modeling is limited by the accuracy of the phenomenological force fields used for protein, DNA and their interactions with each other and with the solvent. Both the AMBER and CHARMM DNA force fields are known to have several drawbacks. Under the conditions when the B-form is to be stable, on time intervals greater than a microsecond, a long DNA molecule loses its duplex structure in the framework of the CHARMM36 force field, and passes into an intermediate form between the B- and A- forms in the framework of the AMBER bsc0 force field \cite{2019-Lyubartsev-AA-FF}. In the 2010s, several torsion angles were edited (primarily $\epsilon/\zeta$) to improve the representation of the BII conformation of phosphates \cite{2016-AMBER-bsc1,2012-CHARMM-for-BII} (AMBER bsc1 and CHARMM36). Recently, in 2021, a correction of the $\alpha/\gamma$ angles has been proposed to more adequately reproduce the Z-DNA in the AMBER force field (OL21) \cite{2021-Amber-for-Z-alpha-gamma}. 

More serious shortcomings for practical purposes are too weak base pairing and too strong base stacking in both the AMBER and CHARMM force fields, with the pairing weaker in the CHARMM force field, and the stacking stronger in the AMBER force field (see, for example, the discussion in Ref.\cite{2022-Amber-CHARMM-comparison-complexes}). Of the nonbonded interactions, DNA-ions interactions (in particular, with Mg$^{2+}$) also do not correspond to the experimental ones. The appropriate corrections for the AMBER force field has resulted in a new force field DES-Amber \cite{2022-new-Amber-for-complexes} with adjusted values for a number of van der Waals parameters and for some atomic partial charges. In the framework of the DES-Amber force field, one mainly uses a 'native' water model, TIP4P-D, since the TIP3P water is known to be a poor solvent for proteins \cite{Piana2015}.

In this regard, let us remind that it was the TIP3P water model that was used to derive the values of partial charges on DNA atoms in the CHARMM force field, and these values have never been revised. In the AMBER force field, a new set of partial atomic charges for DNA and RNA has been devised \cite{2021-W-RESP}. In order to amend excessively strong electrostatic interactions in DNA-protein complexes, a new Tumuc1 force field has been developed \cite{2021-Tumuc1}, with new partial atomic charges and parameters of the valence bonds, angles and torsion angles. Unexpectedly, the Tumuc1 force field simply adapts all the van der Waals parameters from the AMBER force field.

One can do the opposite and weaken too strong DNA-protein interactions by amending their van der Waals, rather than electrostatic, interactions. The CUFIX (or NBFIX) correction \cite{2018-CUFIX-review} adjusts the $\sigma$ parameters of the Lennard-Jones potentials between the atoms of the DNA phosphate group and of some groups of atoms on amino acids. The analogous corrections have been offered also for interactions with ions. It turned out that for each pair of atoms, the parameter $\sigma$ has to be chosen independently of other pairs (combination rules do not apply). The CUFIX correction provides the correct values for the diffusion coefficients of proteins along DNA in both the cases of fast and slow diffusion \cite{Yoo2021}.

\begin{figure*}
\begin{center}
\includegraphics[width=0.99\linewidth]{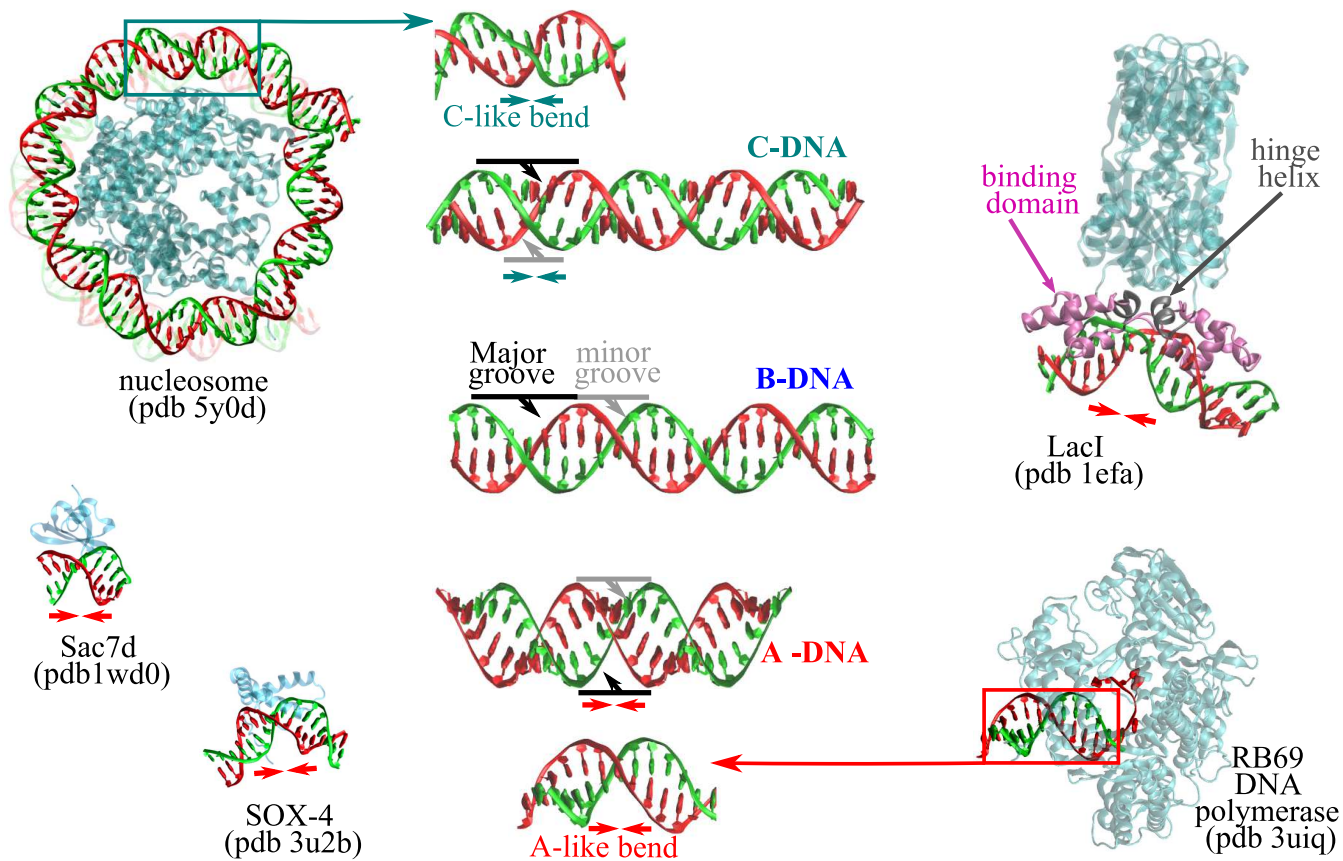}
\caption{Comparison of the ideal C, B, and A-forms of DNA (constructed using the w3DNA 2.0 server \cite{3DNA-site,2019-3DNA}, visualization: VMD program \cite{1996-VMD,VMD-site}).
For the C-DNA, we show the variant with 100\% of BII phosphate conformations. We marked with green and red arrows the local narrowing of the DNA minor and major grooves on the C-like and A-like DNA bends on the nucleosome and in the complex with polymerase, respectively. We also show several more protein complexes with the A-like DNA bends (LacI, SOX-4, and Sac7d).   
\label{fig-C-B-A}
}
\end{center}
\end{figure*}

However, the most difficult task, currently unsolved, is the adequate modeling of the A-form of DNA. DNA takes the A-form locally in complexes with endonucleases, transcription factors, and polymerases \cite{2000-Olson-A-DNA-with-proteins,2010-protein-DNA-recognition-Rohs}. This is a form with a narrow major groove and sugars adopting the north conformation (C3'endo). The geometry of the A-DNA is opposite to that of the C-DNA, in which the minor groove is narrow (see~Figure~\ref{fig-C-B-A}). The signs of the Roll and Slide parameters are also opposite in the A- and C- forms. In the C-DNA, a significant fraction (over 40\%) of phosphates adopts the BII conformation \cite{2000-C-DNA-Levitt}. DNA in the nucleosome has about 20\% of phosphates in the BII conformation (see, for example, the structure 1kx5.pdb \cite{2002-nucleosome-pdb}).

The A- and C- forms of DNA are competitors under conditions with a low dielectric constant of the medium, or a low water activity, or a small available volume. These are the conditions in hydrophobic protein cavities. In vitro, the competition can be demonstrated in experiments when one increases the alcohol proportion in a dilute DNA aqueous solution \cite{1973-Ivanov-CD-A-B-C}. If there is no additional salt in the solution, then, at 78\% of ethanol, the natural DNA (calf thymus) takes the A-form. But, with only 0.1M of NaCl salt, the addition of methanol causes the transition to the C-form. 
In complexes with proteins, DNA can locally take the non-canonical A- and C-forms or be strongly bent \cite{2021-we-PDB}. In these cases, it is important that, for conditions corresponding to the experiment, the force field adequately reproduces the transition to non-canonical conformations of the torsion angles on the DNA sugar-phosphate backbone, including the transition to non-canonical deoxyribose conformations (usually north, C3'endo).

Unfortunately, the transition to the A-form is not adequately modeled by both the AMBER and CHARMM force fields. They are considered to be B- and A-philic, correspondingly \cite{1997-A-B-Amber-ethanol-Cheatham,2018-bad-sugar-for-A-B-in-Amber,2019-A-B-in-SHARMM, 2005-Pastor,2019-Lyubartsev-AA-FF}. Since the balance of interactions in the molecule depends on many factors, changing any of them can easily amend the situation in one particular case. It is more important to find out exactly which parameters of the force fields do not correspond to reality. 

The answer to this question requires a direct comparison of the results of simulations and experiments on the B to C and B to A transitions. In an adequate force field, the DNA molecule is to take both the C- and A- forms (depending on the base sequence) under the experimental conditions. For the test, one has to choose the experiments with the simplest possible conditions, without the interaction of the DNA with complex molecules (alcohol; spermine; drugs; polymers, including DNA: in fibers or in crystals). The simplest system for molecular modeling is a DNA aqueous solution with the addition of salt. We have analyzed (Section~\ref{section-exp-DNA-in-water}) the available experimental data on such transitions for NaCl salt (obtained by Raman and IR spectroscopy \cite{2005-Thomas-Raman-review,2003-library-IR-DNA} and circular dichroism \cite{2009-Kypr-CD-review}).

Based on the analysis, we have formulated the "C-A"{} test for the balance of interactions between the base stacking and the backbone flexibility in DNA force fields. We have tested (Section~\ref{section-sim-DNA-in-water}) the AMBER bsc1 and CHARMM36 force fields and their "hybrids"{}: the fields with potentials for the sugar-phosphate backbone from one force field, and for the base stacking interactions - from another field. For both the AMBER and CHARMM force fields, we have localized the weak points not allowing one to reproduce the experimentally observed DNA behavior. Finally, in the light of the results obtained, we have reviewed (Section~\ref{section-complexes}) the possible influence of the choice of force field on the results of modeling DNA complexes with three different proteins (LacI \cite{2019-DNA-protein-binding-MD-A-DNA-lactose-repressor}, Sox-4 \cite{2021-DNA-protein-binding-MD-rotation}, and Sac7d \cite{Zacharias2019}), where the DNA is bent towards the major groove and locally takes the A-form.
\section{C-DNA and A-DNA in aqueous solution: experiment}
\label{section-exp-DNA-in-water}
\noindent
Circular dichroism (CD) measurements show that, with a large amount of additional salt (4-6M NaCl) in aqueous solution, the natural DNA (calf thymus) takes the C-form (strictly speaking, not completely) \cite{1973-Ivanov-CD-A-B-C}. The regular polymer poly[d(AC)]$\cdot$poly[d(GT)] (poly(AC)) has a CD spectrum very close to that of the calf thymus DNA. Poly(AC) also passes into the C-form upon addition of 4.5M NaCl \cite{1985-Kypr-polyAT-polyAC-polyGC-in-sault}. According to spectral measurements, the polymer poly[d(GC)]$_2$ (poly(GC)) passes into the Z-form \cite{1985-Nishimura-A-DNA-polyG-polyGC}, but, at a low salt concentration (up to 0.2M), poly(GC) is in the classical B-form. Under the same conditions, the polymer poly(dA)$\cdot$poly(dT) (poly(A)) also takes a slightly deformed B-form with some share of sugars in the north conformation (C3'endo) \cite{1986-polyA-polyAT-solution-Nishimura}. The poly(A) double helix has not been observed in any other form under any conditions. A bias toward the A-form is observed in the (double stranded) oligomers d(ATAT)$_2$ (ATAT) and d(GGTATACC)$_2$ (GGTATACC) \cite{1994-Nishimura-ATAT,1987-GGTATACC-Peticolas} (when they are stable: at low temperatures). Only the polymer poly(dG)$\cdot$poly(dC) (poly(G)) demonstrates the complete transition to the A-form upon addition of the 4M NaCl salt \cite{1985-Nishimura-A-DNA-polyG-polyGC}. Lowering the temperature facilitates this transition. Moreover, even in the almost complete absence of the additional salt (0.03M), 30\% of this polymer is in the A-form \cite{1985-Nishimura-A-DNA-polyG-polyGC}. 

Although the picture of the conformational transitions described above seems clear, quantitative estimates can currently be obtained only from IR or Raman spectra. In these spectra, there are definite bands - markers of the A-form \cite{2005-Thomas-Raman-review,2003-library-IR-DNA}. Unfortunately, using the markers known for the C-form, one  cannot make reliable quantitative estimates. As we will see below, the A markers cannot always provide good accuracy as well. The accuracy can be increased by using 2D IR spectra \cite{2007-IR-high-freq-ABZ-theory,2015-2D-IR-DNA-water-interface,2021-phosphate-vibrations-and-water-2D-IR}. However, such data are not available for the required DNA sequences in the required frequency range.

Certain vibrational modes of the DNA nucleotide change their frequency when the sugar conformation changes. Among these modes are, for example, some deformation modes of the pyrimidine ring on bases. Other modes change their frequency when the geometry of the sugar-phosphate backbone changes. When some of DNA nucleotides are in the A-form, and the rest is in the B-form, one may estimate the fraction of the A-form as the ratio of the integral intensity of the A-form marker band to the sum of the intensities of the A- and B- form bands. 

\begin{figure*}
\noindent
\begin{minipage}[c]{\textwidth}

   \begin{minipage}[c]{0.45\textwidth}
    \centering
      \includegraphics[width=0.98\textwidth]{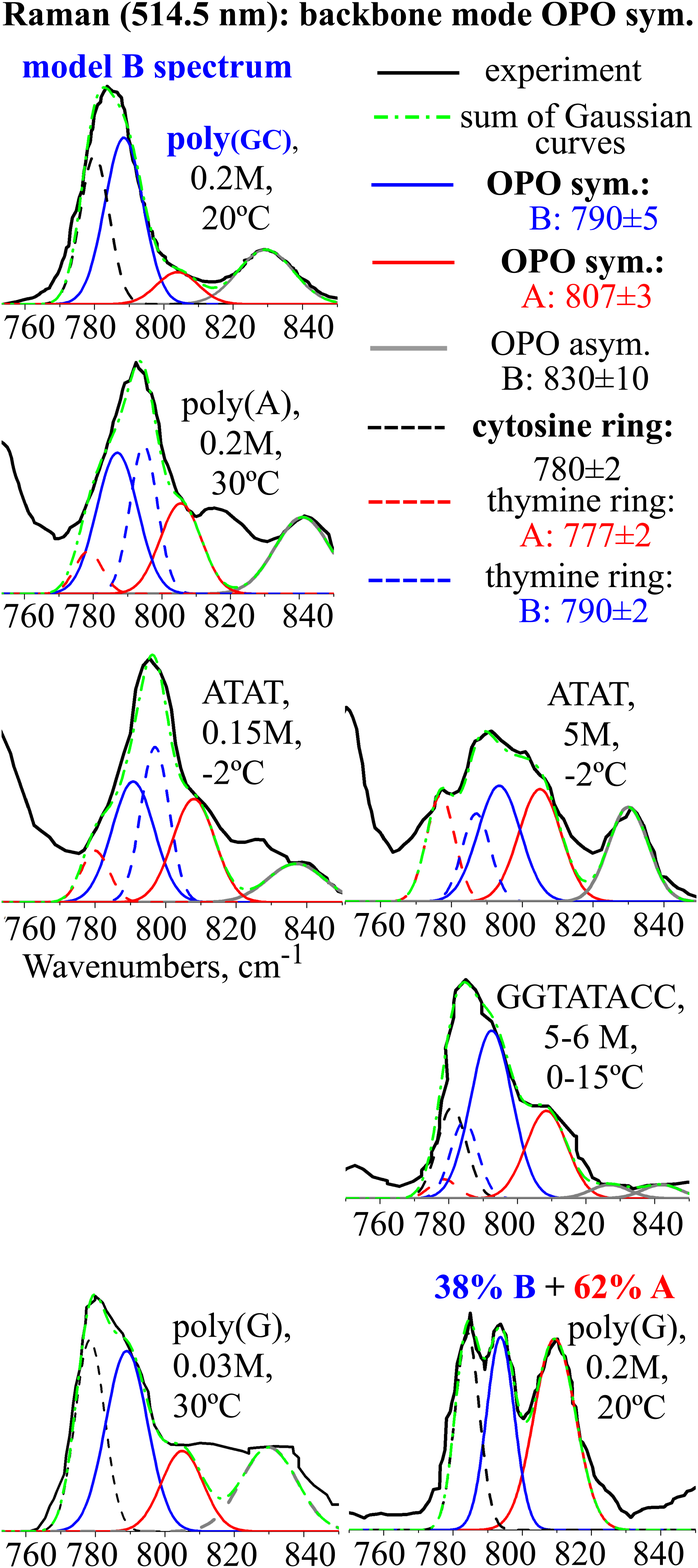}
    \captionof{figure}{Raman spectra in the region of OPOsym (symmetrical stretching in the phosphodiester group) for some regular DNA polymers and oligomers (extracted from \cite{1986-Nishimura-polyG-4M,1985-Nishimura-A-DNA-polyG-polyGC,1994-Nishimura-ATAT}). A conjectured Gaussian peak fitting is done by us.}
  \label{fig-spectra-OPO}
  \end{minipage}
  \hfill  
  \begin{minipage}[c]{0.45\textwidth}
        \captionof{table}{DNA in aqueous NaCl solution. A: A-form according to backbone mode OPOsym. C3: C3'endo sugar pucker; (t): thymine, (g): guanine, (a): adenine; in brackets: wavenumbers for the bands in cm$^{-1}$.}
        \label{table-DNA-in-water}
    \centering
\begin{tabular}{|c|c|c|c|}
\hline 
\hline
DNA                                                                            & NaCl, M & Markers \cite{2005-Thomas-Raman-review} & Form  \\
\hline
\hline

polyGC                                                                                         &0.2       &OPOsym: 17\%A  & \bf{B} \\
20\textdegree C\cite{1985-Nishimura-A-DNA-polyG-polyGC} &               & g(664): 0\%C3  &  \\
\hline   
polyA                                                                                            &  0.2      &OPOsym: 39\%A?               & \bf{B}\rm/$\sim$30\%A\\
30\textdegree C\cite{1986-polyA-polyAT-solution-Nishimura}  &                &  t(777): 22\%C3?             &        \\ 
                                                                                                      &                & [a+t](642): 14\%C3          &       \\
                                                                                                      &                 &\textit{t(1239): 53\%C3}  &       \\
\hline
ATAT\cite{1994-Nishimura-ATAT}                                            & 0.15        &OPOsym: 46\%A?                      & \bf{B}\rm/$\sim$35\%A\\
-2\textdegree C                                                                            &                   & t(777): 25\%C3?                      & \\
                                                                                                     &                  &[a+t](642): 11\%C3                    &   \\
                                                                                                      &                  & \textit{t(1239): 55-62\%C3}    &\\
                                                                                                      & 5             &OPOsym: 49\%A?                      & \bf{A}\rm/\bf{B}\\
                                                                                                       &                    & t(777): 55\%C3?                    & \\
                                                                                                       &                  &[a+t](642): 23\%C3                   &\\                                                                                                                                                                                      
                                                                                                       &                  &\textit{t(1239): 52-61\%C3}   &\\                                                                                                                                                                                                                                                                                 
\hline                                                                                         
GGTA-                                                                               &5-6   & OPOsym: 34\%A? & \bf{B}\rm/$\sim$40\%A\\
TACC\cite{1987-GGTATACC-Peticolas}                        &            & t(777): 20\%C3? & \\
0-15\textdegree C                                                             &            & g(1318): 52\%C3 &  \\
                                                                                          &            &\textit{t(1239): 55\%C3} &\\
\hline 
polyG                                                                                           & 0.01      &OPOsym: 44\%A      &  \bf{B}\rm ?  \\
20\textdegree C\cite{1985-Nishimura-A-DNA-polyG-polyGC}   &              &  g(664): 24\%C3   &  \\
                                                                                                        & 0.2       & OPOsym: 62\%A   &$\sim$60\% \bf{A}      \\
                                                                                                         &             &  g(664): 73\%C3    &\\
\hline
polyG                                                                                  & 0.03      &OPOsym: 31\%A?     & \bf{B}\\
30\textdegree C\cite{1986-Nishimura-polyG-4M}           &               &g(664): 0\%C3         &     \\  
                                                                                           &              & g(1318): 38\%C3     &  \\                                                   
                                                                                            & 4           & OPOsym: 78\%A    &   \bf{A}  \\
                                                                                            &               &g(664): 100\%C3    &   \\
                                                                                            &               &g(1318): 78\%C3      &     \\
10\textdegree C\cite{1986-Nishimura-polyG-4M}            & 1             &OPOsym: 60\%A     & $\sim$60\% \bf{A}\\
                                                                                             &                 & g(664): 65\%C3    & \\
\hline
\hline 
\end{tabular}
    \end{minipage}
 \end{minipage}
\end{figure*}

The transition from the B-DNA to the A-DNA is accompanied by the shift in the frequency of the mode OPOsym (symmetrical stretching vibrations in the phosphodiester group (C-O3'-P-O5'-C)) from 790 to 807~cm$^{-1}$. For DNA molecules consisting only of G:C pairs (for example, poly(G) and poly(GC)), this band can be used - but only at high salt concentrations - to fairly accurately estimate the degree of transition of the backbone to the A-form (see ~Table.~\ref{table-DNA-in-water} and Figure~\ref{fig-spectra-OPO}). Unfortunately, in this region of the spectrum, in addition to this A marker, there is also a band at a frequency of 780~cm$^{-1}$, which corresponds to a certain deformation mode of the cytosine ring. It is possible to separate this band from the OPOsym band only if the salt concentration is high (see~Figure~\ref{fig-spectra-OPO}). In other cases, this marker does not allow quantifying the fraction of the A-form without additional assumptions.

The situation is even worse for oligomers containing A:T pairs. There is a band of thymine similar to that of cytosine at 780~cm$^{-1}$, but the frequency of this band shifts after the B-A transition. It is possible that the intensity of this band also changes, as does the intensity of the OPOsym band. Therefore, the optical spectra (both IR and Raman) do not allow quantifying the fraction of the backbone that has passed into the A-form; only qualitative evaluation is possible. Therefore, for the majority of oligomers and polymers in the Table~\ref{table-DNA-in-water}, we denote the estimates for the OPOsym mode and for the mode of thymine at 777~cm$^{-1}$ with a question mark. In Figure~\ref{fig-spectra-OPO}, we give only one of many possible curve fitting of the experimental spectrum by a sum of Gaussians, it cannot be considered reliable.

Another problem for with A:T pairs is the coincidence of the frequencies (642~cm$^{-1}$) of the pyrimidine ring breathing modes for the adenine and the thymine. It is impossible to separate these bands (and estimate their widths and intensities in different DNA forms) from Raman or IR spectra. Therefore, in the table~\ref{table-DNA-in-water} we present the total estimate for this marker.

The polymer poly(GC) is believed to be in the B-form at low salt concentrations (0-0.2M). In the spectrum of poly(A), there is a noticeable shoulder (at 807~cm$^{-1}$) indicating the presence of north sugars. The spectrum of poly(A) also has a large peak at  820~cm$^{-1}$ (not marked out in Figure~\ref{fig-spectra-OPO}). This band with different heights is present on all the Raman spectra from the DNA samples. In the work \cite{1986-polyA-polyAT-solution-Nishimura}, the authors have even introduced a special designation, $a2$, for a DNA form with certain (non-canonical) torsion angles. The band at  820~cm$^{-1}$ was a marker of this new form. Later, the form $a2$ disappears from the works on IR and Raman spectroscopy of DNA. There is no generally accepted assignment of this band.

The oligomer ATAT exists as a duplex only at low temperatures. From the broadening of the band near 790~cm$^{-1}$, one can definitely say that the addition of salt shifts the conformation towards the A-form, but one cannot give reliable quantitative estimates. The same can be said about the oligomer GGTATACC.

The most A-philic polymer is poly(G). Even at the lowest salt concentration (0.03M) and a rather high temperature of 30\textdegree C, the band OPOsym has a shoulder corresponding to the A-form. At room temperature and a salt concentration of 0.2 M, the backbone of this polymer is 62\% in the A-form.

Let us compare the fractions of north sugar conformations calculated using different markers. The high-frequency thymine deformation mode at 1239~cm$^{-1}$ gives approximately the same estimate of $\sim$55\% for all the cases presented in Table \ref{table-DNA-in-water}. However, judging by the backbone marker OPOsym, the polymer poly(A) is in the (slightly deformed) B-form, and the oligomer ATAT undoubtedly contains a significant fraction of the A-form at the high salt concentration. In the works cited in the Table~\ref{table-DNA-in-water}, there is only one case when the estimate by this marker differs from $\sim$55\%. This is a crystal GGTATACC in the A-form with t(1239):$\sim$67\%. A similar band for the guanine deformation mode (breathing of the pyrimidine ring) at 1318~cm$^{-1}$ behaves quite differently. For the polymer poly(G), the estimate g(1318) practically repeats the estimate by the backbone marker OPOsym (in the cases when this estimate can be considered reliable). In addition, the estimate by the marker [a+t](642) is always very low. Judging by these results, the use of the markers t(1239) and [a+t](642) cannot be considered justified without further analysis of their validity.

The practical conclusion from the experimental data reviewed above is as follows. Two DNA sequences, poly(AC) and poly(G), behave oppositely in aqueous NaCl solution at a high salt concentration. The first passes into the C-form (the minor groove narrows), the second into the A-form (the major groove narrows). Strictly speaking, one may limit oneself to these two sequences to test a DNA force field for the balance between the flexibility of phosphates and sugar puckers. Other oligomers can serve as reference points. The polymers poly(GC) and poly(A) at low salt concentrations are the variants of the B-form (the classical form and a form with a shift to the A-DNA). For the A-philic oligomers ATAT and GGTATACC, the fraction of the A-form is to increase with the addition of salt.
\section{C-DNA and A-DNA in aqueous solution: modeling}
\label{section-sim-DNA-in-water}
\noindent
We have simulated all the six DNA sequences discussed in the previous section: poly(GC) and poly(A) (20 bp duplexes); ATAT and GTATACC; poly(AC) and poly(G) (20 bp duplexes) at different salt concentrations (see~Table~\ref{table-simulations-summary}). We have tested four force fields: AMBER (bsc1), CHARMM (36) and two hybrid fields, in which the sugar-phosphate backbone was modeled in the framework of one field, and the base stacking - in the framework of the other force field (for details, see Section~\ref{section-MD-himeras}). In all the cases, we used the same water and ion models (TIP4P/long \cite{2004-TIP4P-Ew} water, Joung-Cheatham ions \cite{2008-Young-Cheatham-ions}).

\begin{figure}
\begin{center}
\includegraphics[width=0.75\linewidth]{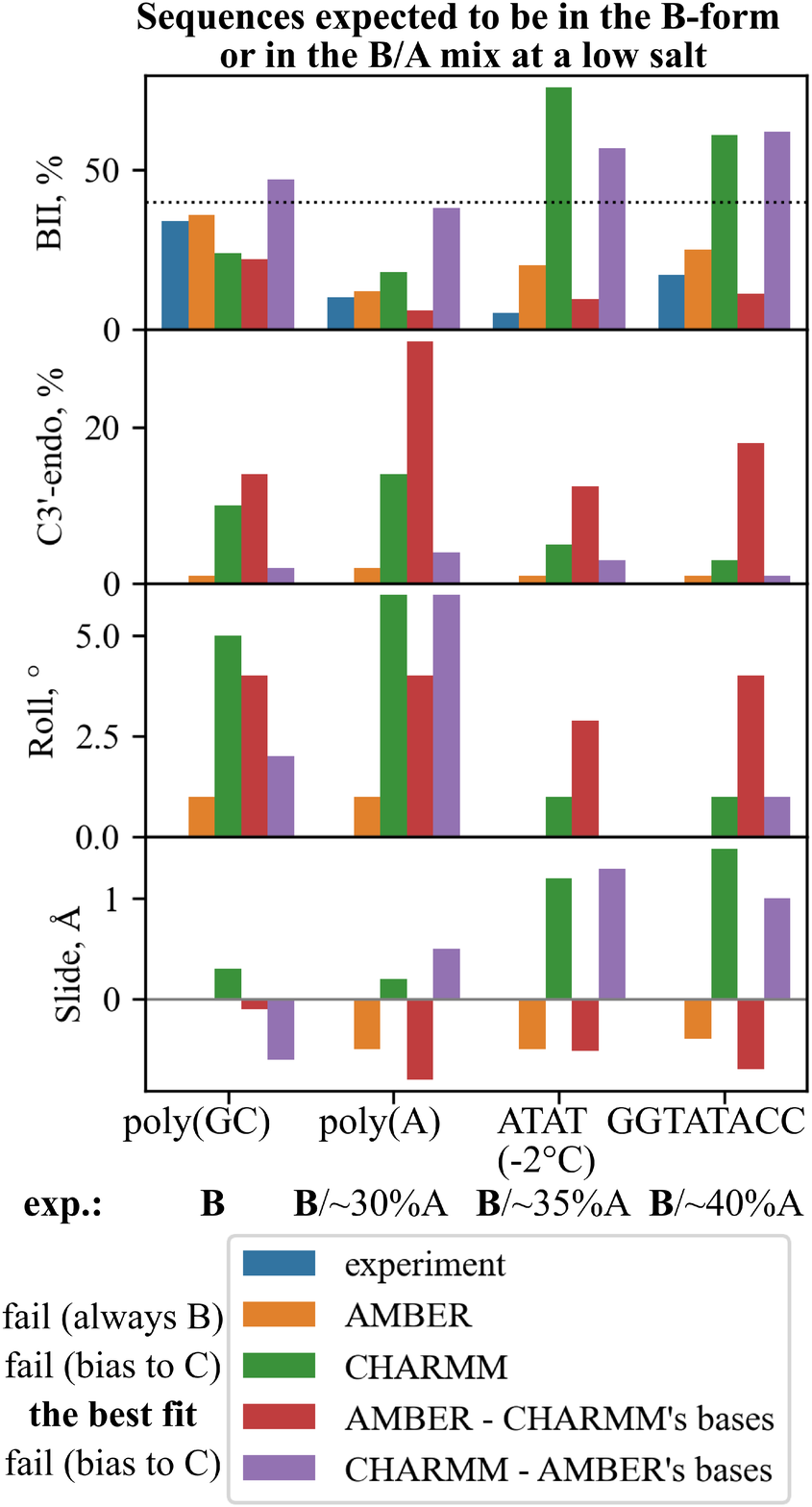}
\caption{Graphical representation of Table~\ref{table-no-transition-chimeras-in-water-simulations}. The experimental data on the fraction of the A-form are taken from Table~\ref{table-DNA-in-water}.
\label{fig-histogram-no-transition}
}
\end{center}
\end{figure}

Table \ref{table-no-transition-chimeras-in-water-simulations} and figure~\ref{fig-histogram-no-transition} show the simulation results for the first four oligomers at a low salt concentration. The first defect of all the tested force fields is that none of them (except for the hybrid 'AMBER - CHARMM's bases' in the case of the oligomer ATAT) showed any dependence on salt concentration for the oligomers ATAT and GGTATACC. This contradicts the experiment (higher salt concentration - larger fraction of the A-form). All the system parameters at the high salt proved out to be the same as at the low salt, so they are not listed in Table~\ref{table-no-transition-chimeras-in-water-simulations} and in Figure~\ref{fig-histogram-no-transition}.
\begin{table*}
\caption{Conformations of the DNA oligomers for which no dependence on the salt concentration was found in the simulations (except for ATAT for the force field 'AMBER -CHARMM's bases'). Initially, all the oligomers were in the B-form. Unless otherwise indicated, the transition to the final conformation occurred within 5~ns after the system preparation (see~Section~\ref{section-MD-himeras}). Temperature: 300K (unless otherwise stated), additional salt: $\sim$0.15M NaCl, water: TIP4P/long \cite{2004-TIP4P-Ew}; ions: Joung-Cheatham \cite{2008-Young-Cheatham-ions}. For each DNA sequence, the average experimental value of the fraction of BII phosphate conformations in solutions is given (by $^{31}$P NMR chemical shift \cite{2009-chem-shift-Hartman}). C3: north sugar pucker (C3'endo) of thymine (t), cytosine (c), guanine (g) or adenine (a). The units of the parameters Roll and Slide are degrees and angstroms, respectively.}
\label{table-no-transition-chimeras-in-water-simulations}
\begin{tabular}{|c|c|c|c|c|}
\hline 
\hline
DNA,     & AMBER                        & CHARMM                                     & AMBER -                       & CHARMM -        \\ 
exp. \%BII &------------------------------ &------------------------------&---- CHARMM's bp ----&----- AMBER's bp ----- \\ 
\hline
\hline
polyGC   & \bf{B}                   &\bf{B}                          &\bf{B}                                 & \bf{B}   \\ 
34\% BII &g:0($\pm$1)\%C3  &g:3($\pm$4)\%C3       &g:4($\pm$4)\%C3               &g:0.1($\pm$0.6)\%C3 \\
               &c:2($\pm$3)\%C3   &c:16($\pm$9)\%C3      &c:23($\pm$14)\%C3          &c:3($\pm$4)\%C3\\ 
               & 36($\pm$5)\% BII      & (24$\pm$7)\% BII    & (22$\pm$7)\% BII         &(47$\pm$9)\% BII \\
                &Roll=1                    &Roll=5                            &Roll=4                             &Roll=2          \\
                & Slide=0                  & Slide=0.3                      & Slide=-0.1                       &Slide=-0.6     \\
\hline

polyA      &  \bf{B}                       &\bf{B}                                   &\bf{B}                                        &\bf{C} \rm/ \bf{B}      \\ 
10\% BII &2($\pm$2)\%C3           &a:4($\pm$4)\%C3                &a:25($\pm$14)\%C3                 &a:1($\pm$2)\%C3 \\
                &                                    &t:23($\pm$8)\%C3                &t:36($\pm$14)\%C3                  &t:6($\pm$5)\%C3\\ 
                & 12($\pm$4)\% BII      & (18$\pm$7)\% BII              & (6$\pm$3)\% BII                      &(38$\pm$9)\% BII \\
                &Roll=1                        &Roll=6                                  &Roll=4                                        &Roll=6                     \\ 
                &Slide=-0.5                    &Slide=0.2                             &Slide=-0.8                                  &Slide=0.5                \\
\hline
\hline
ATAT                    &  \bf{B}                       & \bf{C},                          & \bf{B}                               &  \bf{C},                 \\ 
(-2\textdegree C)  &                                    & terminal H bonds          &                                          & terminal H bonds \\
5\% BII                &                                     &broken after sev. ns      &                                         &       off and on \\
                             &1($\pm$5)\% C3         & 5($\pm$9)\% C3           & 12($\pm$11)\% C3         &  3($\pm$6)\% C3\\ 
                             & 20($\pm$12)\% BII    & 76($\pm$17)\% BII      &  7($\pm$10)\% BII       &  57($\pm$19)\% BII   \\
                             & Roll=0                         & Roll=1,                          &  Roll=3                         &  Roll=0             \\ 
                             &  Slide=-0.5                   &Slide=1.2                       &Slide=-0.4                      & Slide=1.3         \\
\hline
GGTA-          &  \bf{B}                    &     \bf{C}                         &\bf{B}                  &     \bf{C}                      \\  
TACC           &1($\pm$3)\%C3       & 3($\pm$5)\%C3             &a,t:19($\pm$19)\%C3                & 1($\pm$3)\%C3\\ 
17\% BII        &                                &                                         &g:30($\pm$29)\%C3                 &\\
                       &                                 &                                        &c:4($\pm$10)\%C3                   &\\
                       & 25($\pm$8)\%BII   & 61($\pm$17)\%BII         &11($\pm$8)\%BII                     &  62($\pm$18)\%BII   \\
                       & Roll=0                    & Roll=1                             & Roll=4                                     & Roll=1                      \\  
                       & Slide=-0.4              &  Slide=1.5                        & Slide=-0.7                                &  Slide=1                  \\            
\hline 
\hline
\end{tabular} 
\end{table*}

In the AMBER force field, at any salt concentration, these four oligomers retain the B-form, the oligomer ATAT exhibiting an excess of BII phosphate conformations (at low temperature). In the CHARMM force field, the oligomers GGTATACC and ATAT are in the C-form even at low salt. The  bias towards the C-form is the defect of the sugar-phosphate backbone of the CHARMM force field, because this bias is absent in the hybrid force field 'AMBER~-~CHARMM's bases', but is present in the force field 'CHARMM~-~AMBER's bases'. 

On the other hand, in the CHARMM force field the base stacking is A-philic. Indeed, in the hybrid force field 'AMBER~-~CHARMM's bases', the amount of north sugars is much higher for all four oligomers and is in the best agreement with the experiment. Moreover, this force field reproduces an increase in the fraction of north sugars with an increase in the salt concentration for the oligomer ATAT. At a salt concentration of 5 M, the replicas (we had four of them) pass from one form to another, being in the B-form more often. So one can say that the result is a mixture of the B- and A-DNA. On average, the fraction of north sugars is 24($\pm 16$)\% (at 5($\pm 9$)\% BII phosphate conformations; Roll=5\textdegree, Slide=-0.7\AA). 

\begin{figure*}
\begin{center}
\includegraphics[width=0.75\linewidth]{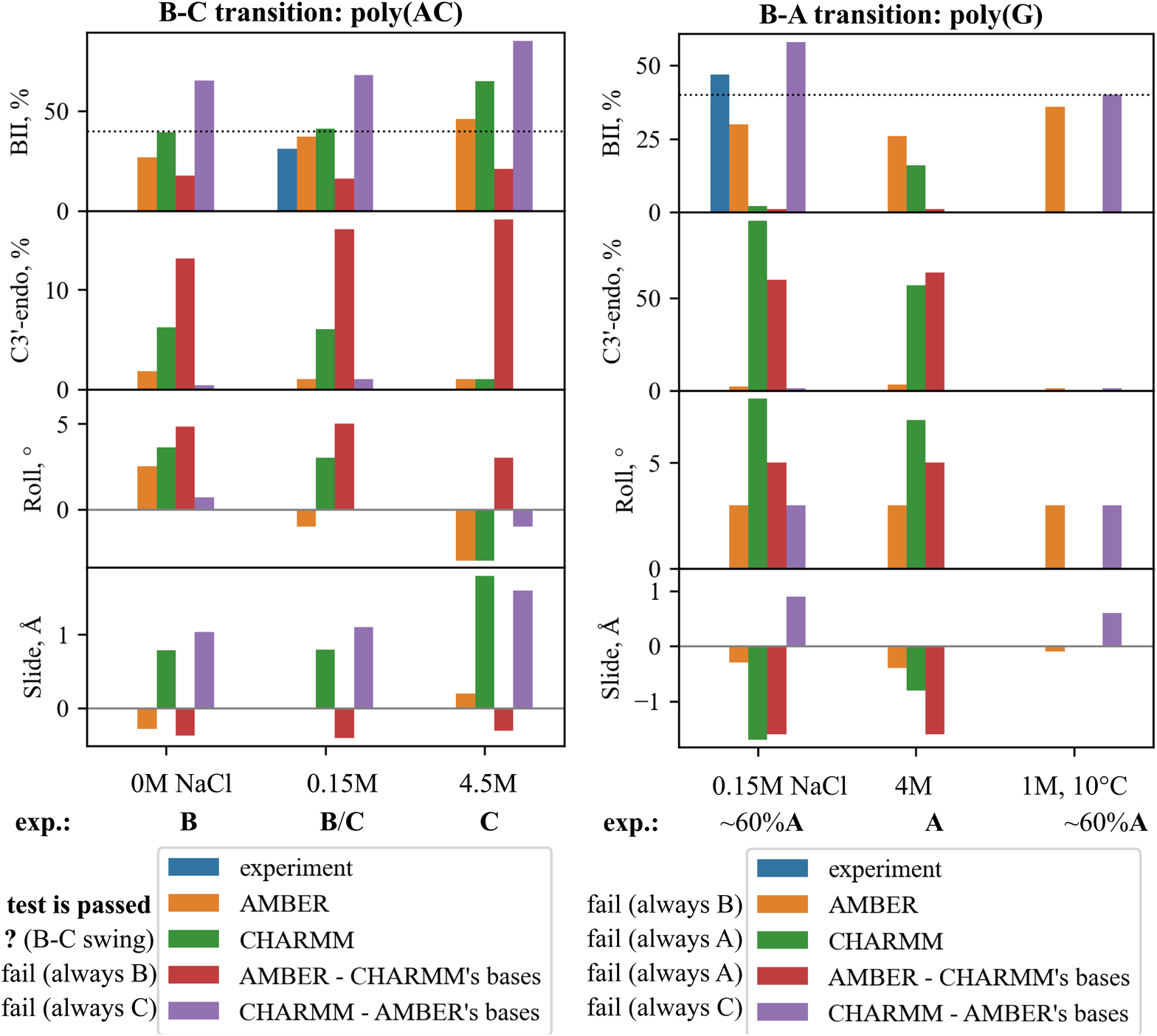}
\caption{
Graphical representation of Tables \ref{table-B-C-chimeras-in-water-simulations} and \ref{table-B-A-chimeras-in-water-simulations}. The experimental data on the fraction of the A-form are taken from Table~\ref{table-DNA-in-water}.
\label{fig-histogram-polyAC-polyG}
}
\end{center}
\end{figure*}

For the polymers poly(AC) and poly(G), a salt dependence does take place (see Tables \ref{table-B-C-chimeras-in-water-simulations} and \ref{table-B-A-chimeras-in-water-simulations}, and Figure \ref{fig-histogram-polyAC-polyG}). The AMBER force field reproduces the transition of the polymer poly(AC) from the B- to the C- form with the addition of salt, but does not reproduce the bias of the polymer poly(G) to the A-form, nor the complete transition to the A-form at high salt concentration. In the AMBER force field, this polymer retains the B-form at any salt. Moreover, a decrease in temperature increases the bias of poly(G) in the opposite direction: towards the C-form. 
\begin{table*}
\caption{Dependence of the conformation of the polymer poly(AC) on NaCl concentration in different force fields (in the experiment, one observes the transition from the B- to the C- form with increasing salt). Simulation parameters and notation are the same as in Table~\ref{table-no-transition-chimeras-in-water-simulations}.}
\label{table-B-C-chimeras-in-water-simulations}
\begin{tabular}{|c|c|c|c|c|}
\hline 
\hline
DNA,        & AMBER                                                   & CHARMM                    & AMBER -               & CHARMM -        \\ 
exp. \%BII &------------------------------ &------------------------------&---- CHARMM's bp ----&----- AMBER's bp ----- \\ 
\hline
\hline
polyAC   &  \bf{B}                                                    & alternate B-C              &  \bf{B}                        & \bf{C}           \\ 
0M          & after 20ns failed                                        & and C-B trans.           &                                    &after 15ns relax. \\ 
               & B-C trans. attempt                                    &                                    &                                  &                            \\ 
                &2($\pm$2)\%C3                                         &6($\pm$5)\%C3        &13($\pm$7)\%C3         &0($\pm$1)\%C3\\
                & 27($\pm$5)\% BII                                    & (40$\pm$10)\% BII  & (18$\pm$5)\% BII      &(65$\pm$10)\% BII \\
                &Roll=3                                                       &Roll=4                        &Roll=5                         &Roll=1        \\
                &Slide=-0.3                                                  &Slide=0.8                   & Slide=-0.4                   & Slide=1 \\
                \hline
polyAC   &  \bf{C}\rm/$\sim$3ns jumps to \bf{B} & alternate B-C              &  \bf{B}                       & \bf{C}           \\ 
0.1M     & after 30ns in B,                                        & and C-B trans.           &                                      &after 15ns in B, \\ 
31\% BII & 5ns B-C trans.                                        &                                    &                                  & 5ns B-C trans.\\ 
                &1($\pm$1)\%C3                                         &6($\pm$6)\%C3        &16($\pm$6)\%C3       &1($\pm$1)\%C3\\
                & 37($\pm$5)\% BII                                    & (41$\pm$12)\% BII  & (16$\pm$5)\% BII    &(68$\pm$10)\% BII \\
                &Roll=-1                                                     &Roll=3                        &Roll=5                          &Roll=0           \\
                &Slide=0                                                      &Slide=0.8                   &Slide=-0.4                    &Slide=1.1 \\
\hline
polyAC   &  \bf{C}                                                      &  \bf{C}                      &  \bf{B}                       & \bf{C}           \\ 
4.5M       & after 13ns B-C trans.,                                 &after 45ns                    &                                   &after 20ns  \\ 
                & 22ns relax.                                                 & B-C trans.                   &                                   & B-C trans.      \\ 
                &1($\pm$1)\%C3                                         &1($\pm$1)\%C3        &17($\pm$5)\%C3         &0($\pm$1)\%C3\\
                & 46($\pm$2)\% BII                                    & (65$\pm$7)\% BII    & (21$\pm$6)\% BII      &(85$\pm$8)\% BII \\
                &Roll=-3                                                     &Roll=-3                      &Roll=3                          &Roll=-1                \\ 
                &  Slide=0.2                                                 &  Slide=1.8                 & Slide=-0.3                    &  Slide=1.6  \\              
\hline
\hline 
\end{tabular} 
\end{table*}
\begin{table*}
\caption{Dependence of the conformation of the polymer poly(G) on NaCl concentration in different force fields (in the experiment, one observes the full transition to the A-form with increasing salt). Simulation parameters and notation are the same as in Table~\ref{table-no-transition-chimeras-in-water-simulations}.}
\label{table-B-A-chimeras-in-water-simulations}
\begin{tabular}{|c|c|c|c|c|}
\hline 
\hline
DNA,        & AMBER                           & CHARMM                 & AMBER -               & CHARMM -        \\ 
exp. \%BII &------------------------------ &------------------------------&---- CHARMM's bp ----&----- AMBER's bp ----- \\ 
\hline
\hline
polyG       &   \bf{B}                            &  \bf{A}                              &  \bf{A}                                &  \bf{C}  \\ 
0.15M       &                                         & after 30ns B-A trans.         &                                             &           \\ 
47\% BII  & 2($\pm$2)\%C3              &g:93($\pm$6)\%C3            &g:87($\pm$6)\%C3              &1($\pm$1)\%C3\\ 
                 &                                         &c:82($\pm$7)\%C3            &c:34($\pm$16)\%C3             &                            \\ 
                &30($\pm$6)\%BII            &3($\pm$3)\% BII                &1($\pm$2)\%BII                  &58($\pm$10)\%BII  \\
                 &Roll=3                            & Roll=9                                 &Roll=5                                   &Roll=3  \\
                 &Slide=-0.3                       &Slide=-1.6                            &Slide=-1.6                             &Slide=0.9  \\
\hline                 
polyG       &   \bf{B}                            &  \bf{A} \rm/ \bf{B} mix       &  \bf{A}                                &   \\ 
4M           &                                         & after 35ns trans.                     &                                             &           \\ 
                & 3($\pm$3)\%C3             &g:59($\pm$11)\%C3              &g:86($\pm$6)\%C3            &\\ 
                &                                         &c:54($\pm$13)\%C3             &c:42($\pm$14)\%C3            &\\ 
                 &26($\pm$6)\%BII            &16($\pm$7)\% BII                &1($\pm$1)\% BII                  &  \\
                 &Roll=3                              & Roll=7                                 &Roll=5                                    &  \\ 
                 & Slide=-0.4                        & Slide=-0.8                            & Slide=-1.6                            &\\                   
\hline
polyG                  &   \bf{B}  \rm/ \bf{C}     &                               &                                             &  \bf{B} \rm/ \bf{C}  \\ 
1M,                     & 1($\pm$2)\%C3             &                               &                                              &1($\pm$1)\%C3\\ 
10\textdegree C  &36($\pm$5)\%BII          &                               &                                              &40($\pm$6)\%BII  \\
                           &Roll=3                            &                               &                                                &Roll=3                      \\ 
                           & Slide=-0.1                      &                               &                                                &Slide=0.6                 \\                      
\hline 
\hline
\end{tabular} 
\end{table*}

In the CHARMM force field, the polymer poly(AC) is never in the B-form, even with zero additional salt. There are successive transitions from B to C and vice versa. At the high salt concentration, a complete transition to the C-form occurs. The polymer poly(G) is completely in the A-form already at a salt of 0.15M (which does not correspond to the experiment). As the salt concentration increases, the fraction of the A-form decreases (also in contrast to the experiment), while the fraction of BII conformations increases.

In the force field 'AMBER~-~CHARMM's bases', the  polymer poly(G) is in the A-form, regardless of the amount of salt, which confirms the bias of the base stacking (rather than the backbone) in the CHARMM force field to the A-form. In the CHARMM force field at the high salt, the narrowing of the minor groove and the partial transition to the B- and then to the C-form (the fraction of phosphates in the BII conformation increases from 3 to 16\%) is provided by the sugar-phosphate backbone. Indeed, in the force field 'CHARMM~-~AMBER's bases' (in which the AMBER bases do not tend to A-like base stacking), even at the low salt, the polymer poly(G) is in the C-form.

The bias of the DNA backbone in the CHARMM force field toward the C-form may be provided by the interaction of charges on the backbone with ions in solution. At the high salt, a large amount of sodium ions appear in the minor groove of poly(G) near the atom O4' of the sugar ring (direct contact) and near the atom N2 of guanine. Perhaps the reason is that the  atom O4' in the CHARMM force field has a charge of -0.5 versus -0.3691 in the AMBER force field. These additional sodium ions in the minor groove, attracting phosphate groups, seem to provide narrowing of the minor groove and the shift in the balance toward the C-form.

In summary, the base stacking of the AMBER force field is excessively B-philic with a slight bias towards the C-form. Contrary to experiment, there is no A-DNA for the polymer poly(G) in this force field. However, it reproduces the transition to the C-form for the polymer poly(AC). In the CHARMM field, the base stacking provides the bias to the A-form for poly(G), while the backbone is excessively C-philic. It can be said that the B-DNA in the CHARMM force field is the result of a fluid balance between the shift of the base stacking towards the A-form and the backbone - towards the C-form. In contrast, the B-DNA in the AMBER force field is a stable inflexible construct due to the B-philic excessively strong base stacking.
\section{DNA force field and modeling of DNA-protein complexes}
\label{section-complexes}
\noindent
Let us consider some simulation results on the formation of DNA-protein complexes in the case of a strong A-like DNA bending (towards the major groove, see~Figure~\ref{fig-C-B-A}). If the DNA molecule, in the framework of the applied force field, is not able to adopt the required conformation under experimental conditions, then the complex may not form, even in the case of excessively strong DNA-protein interactions. On the other hand, the complex may not form for other reasons, including technical ones related to other shortcomings of the force field. But the researcher may erroneously conclude that the binding mechanism is a conformational selection (a protein binds only to a certain non-canonical form of DNA), while in reality an induced fit takes place (a protein induces the needed change in DNA conformation during or after binding, see review \cite {2015-DNA-protein-MD-binding-bending-folding-VanDerVaart}).

Let us compare the results of modeling of the DNA/LacI \cite{2019-DNA-protein-binding-MD-A-DNA-lactose-repressor}, DNA/SOX-4 \cite{2021-DNA-protein-binding-MD-rotation} and DNA/Sac7d \cite{Zacharias2019} complexes. All the authors use the AMBER force field (variants bsc1 in Ref.\cite{2019-DNA-protein-binding-MD-A-DNA-lactose-repressor} and bsc0 in Ref.\cite{2021-DNA-protein-binding-MD-rotation,Zacharias2019}) with the TIP3P water. The bsc1 \cite{2016-AMBER-bsc1} modification increases the fraction of BII phosphate conformations (underestimated in bsc0) and facilitates the transition to the C-form.

In the experiment, in the process of the complex formation, all three the proteins LacI, SOX-4, and Sac7d bind to the minor groove and, expanding it, form an A-like bend towards the major groove (see Figure \ref{fig-C-B-A}). DNA locally takes the A-form. In the case of the protein LacI, the binding (and presumably the target recognition) also occurs in the major groove.

In the modeling, the protein Sac7d (albeit at an elevated temperature of 338K) quickly (within tenths of a nanosecond) penetrates the DNA minor groove, moves along it, and induces a bend on the target, showing classical induced fit within a time interval less than a microsecond. The protein SOX-4 also demonstrates fairly fast (within 5 nanoseconds at 300K) non-specific binding. However, for this protein at a lower temperature - and, perhaps, due to too strong DNA-protein interactions in the AMBER force field - the transition from the nonspecific to specific binding seems impossible to observe by ordinary molecular dynamics. The authors do not do this and study the specific binding mechanism with the use of the metadynamics method.

It should be noted that, in most works on DNA-protein complex formation, the authors use enhanced sampling techniques: steered molecular dynamics, metadynamics, umbrella sampling, or weighted histogram analysis method. These approaches make it possible to build free energy profiles and compare the obtained energy of complex formation with experiment. But one can not follow the process of complex formation in real time. In case of insufficient number or poor choice of reaction coordinates, one may face a difficulty in finding the real pass to the global minimum. The discrepancies may also result from an inadequate balance of interactions in other parts of the system (for example, incorrect DNA-protein binding energy, inappropriate water model, and so on, see Section~\ref{section-Intro}).

In the case of the protein LacI, after a microsecond ordinary MD calculation, the authors failed to obtain the native state of the DNA-protein complex with an A-like bend. Contrary to the proteins SOX-4 and Sac7d, the transition to this state is impeded by a steric clash between LacI and the straight DNA. Because of the known B-philicity of the AMBER force field, the authors hypothesized that the DNA molecule is to be already bent before the complex formation (conformational selection). However,  there are two other factors (specifics of the complex formation and the features of the force field not related to DNA) that could contribute to the negative result of the simulation.

There are experimental data \cite{2004-lacI-binding-Science} that two “hinge” alpha-helices, bound with the DNA minor groove in the LacI/DNA complex, are disordered in the free and non-specifically bound protein. More exactly, these data refer to a truncated protein, the "headpiece", consisting of the hinge helices and the binding domains (see Figure~\ref{fig-C-B-A}). But, since the affinity of the headpiece for DNA is the same as of the whole LacI protein, the headpiece is believed to be an adequate model system for binding of the whole protein. In the simulations by Ref. \cite{2019-DNA-protein-binding-MD-A-DNA-lactose-repressor}, the hinge helices in the free (whole) protein are almost ordered. In the CHARMM force field (also with the TIP3P water), isolated hinge helices in solution are disordered \cite{2020-disorder-in-hinge}. But they are ordered as a part of the free LacI headpiece \cite{2020-disorder-in-hinge}, contrary to experiment. Such an excessive stability of the hinge helices may result from using the TIP3P water model which is a poor solvent for proteins \cite{Piana2015}. The same may be true also for the modeling of the whole LacI protein in Ref.\cite{2019-DNA-protein-binding-MD-A-DNA-lactose-repressor}. 

The ordered state of the hinge helices may not facilitate, but, on the contrary, hinder the formation of the complex. Indeed, the folded hinge helices cannot penetrate the minor groove at the early stages of binding. So they cannot destroy the spine of hydration \cite{2017-chiral-spine-of-hydration} in the minor groove, which stabilizes the B-form and prevents binding as well as bending. Without additional research, it is impossible to say what exactly - B-philic DNA of the AMBER force field, inappropriate water model, both together, or something else in addition (as a high barrier for the binding) - caused the negative result of modeling in Ref. \cite{2019-DNA-protein-binding-MD-A-DNA-lactose-repressor}. 

In the cases of SOX-4 and Sac7d, such a complicated rearrangement of the protein is not needed for the complex formation. Therefore, excessively strong DNA-protein interactions may outweigh the DNA rigidity in the AMBER force field. As a result, the protein Sac7d easily induces the required DNA bend. And, for the native DNA/SOX-4 complex, there is a corresponding (global) minimum and a not very high barrier to pass from the (local) minimum for the non-specifically bound complex.

Thus, when analyzing the results of modeling the formation of DNA-protein complexes, it is necessary to take into account both the specifics of the possible mechanism of binding of the particular protein to DNA, and the features of the force fields used. Unfortunately, the choice of the force field is almost never substantiated in the corresponding works. In the case of modeling the complex with the protein LacI, it would be better to choose the CHARMM force field for the DNA. In it, the DNA duplex is more flexible and, for sequences with a large number of G:C pairs, more A-philic. This choice would allow eliminating at least one of the uncertainty factors in the analysis of the obtained simulation results.
\section{Conclusion}
\noindent
The DNA duplex in the DNA-protein formation process often locally bends (nucleosome, complexes with polymerases) and passes into the  C (narrow minor groove, negative Roll, positive Slide) or A (narrow major groove, positive Roll, negative Slide) forms. The B-C transition is accompanied by the switch of a significant fraction (more than $\sim$40\%) of phosphates from the BI to the BII conformation, and the B-A transition is accompanied by the switch of sugars from the south (C2’endo) to the north (C3’endo) conformation. To study the formation of protein complexes with bent DNA by molecular dynamics modeling, the DNA force field is needed to adequately reproduce both the conformational transitions, which requires appropriate testing.

In the present work, we have analyzed the available experimental data on the B-C and B-A transitions under the conditions most conveniently implemented in molecular dynamics modeling: in an aqueous NaCl solution. Unfortunately, from the currently available spectroscopic data (Raman and IR spectroscopy), one cannot extract in most cases reliable quantitative estimates of the fraction of north sugars or of the fraction of the DNA backbone in the A conformation. For the B-C transition,  only the circular dichroism data are reliable. And it is also currently impossible to extract the exact fraction of the BII conformations of phosphates from them. However, there are DNA sequences that can be used to test DNA force fields.

We have selected six DNA oligomers that respond differently to an increase in salt concentration. At a low salt concentration, two polymers, poly(GC) and poly(A), take the B-form, classical and slightly deformed with some fraction of north sugars, respectively. The oligomers ATAT (at a very low temperature, when it is stable) and GGTATACC (also at low temperature) have an appreciable bias towards the A-form. The fraction of the A-form increases with increasing salt concentration. At a high salt concentration of 4-5M, the polymers poly(AC) and poly(G) take the C- and A- forms, respectively.

Modeling these six DNA duplexes at the two (low and high) salt concentrations is a good test for the balance between the base stacking and the backbone (phosphate groups and deoxyriboses) conformational mobility in DNA force fields. We have tested the two most commonly used force fields AMBER bsc1 and CHARMM36. To find out how exactly the balance is upset in each case, we also had to test the hybrid force fields, with backbone interactions taken from one of the force fields, and the base stacking - from another.

We showed that the "B-philicity"{} of the AMBER force field is a consequence of the "B-philicity"{} of its base stacking. The stacking is excessively strong, and it does not allow the polymer poly(G) to pass to the A-form. The transition to the C-form for the polymer poly(AC) is possible in the framework of the AMBER force field. In the CHARMM force field, the balance between the interactions in the DNA duplex is fundamentally different. Namely, the B-form is the result of an fragile balance between the A-philic base stacking (especially for G:C base pairs) and the C-philic backbone. This equilibrium is strongly shifted towards the C-form with the predominance of A:T pairs in the DNA sequence, and towards the A-form with the predominance of G:C pairs. Therefore, in the CHARMM force field, the polymer poly(G) is always in the A-form, and the fraction of sugars in the north conformation for poly(GC) and poly(A) approximately coincide (in the experiment, poly(A) is more A-philic). The question on the balance between the base stacking and the backbone conformational flexibility in the real DNA duplex remains open.

In both the force fields, the interactions between the DNA and the ions make much smaller contribution to the overall balance of interactions in the system as compared to the experiment. In simulation of the oligomers ATAT and GGTATACC the experimental dependence on salt concentration is not observed. The influence of ions in both the force fields is especially weak under the conditions of the B-A transition. The fraction of north sugars in poly(G) does not depend on the salt concentration in the framework of neither force field (except for the wrong dependence in the CHARMM force field); while the B-C transition for poly(AC) is reproduced in the AMBER force field, and almost reproduced in the CHARMM force field. Due to the "dynamic" {} nature of the B-form in the CHARMM force field, at low salt concentrations the polymer poly(AC) incessantly passes between the B- and C- forms, the C-form having too large percentage of BII phosphate conformations.

Thus, we have revealed a discrepancy in the DNA conformational flexibility (transitions to the C- and A- forms) between modeling in the framework of the AMBER and CHARMM force fields and the experimental data. A researcher choosing a force field might want to take into account the revealed features of the DNA duplex in these two commonly used force fields, along with the specifics of the possible mechanism of the complex formation, especially if the formation requires A-like bending of the DNA duplex. For example, in the case of the LacI/DNA complex, the CHARMM force field would be a better choice than the AMBER force field. In the CHARMM force field, the DNA duplex is more flexible and, with the predominance of G:C pairs, more A-philic. However, the reliability of the results can only be ensured by a force field that can withstand the C-A test.
\section{Methods: details of MD calculations}
\label{section-MD-himeras}
\noindent
We simulated DNA oligomers in water with ions using the LAMMPS \cite{LAMMPS-site,LAMMPS-2022} package in the framework of the force fields AMBER bsc1 \cite{2016-AMBER-bsc1} and CHARMM36 \cite{2012-CHARMM-for-BII} and their hybrids, four force fields in total. In the hybrid force fields, the bonded interactions belonged to the first force field (for example, AMBER in the force field ‘AMBER~-~CHARMM’s bases’). The non-bonded interactions in the sugar-phosphate backbone belonged to the same force field, while the van der Waals parameters and partial electric charges on the bases were replaced by their counterparts from the other force field (CHARMM in the force field 'AMBER~-~CHARMM's bases'). Thus, the base stacking (van der Waals and Coulomb interactions between bases) was modeled in the framework of the other force field. We have compensated the difference in total charges on the bases (they are neutral in the CHARMM and negative in the AMBER) by a change in the charge on the hydrogen atom H1'. In such a splicing of the force fields, the effective potential of base rotation around the glycosidic bond $\chi$ turns out to be hybrid. Indeed, the potential depends on hybrid van der Waals and Coulomb interactions between the atoms on the sugar ring (with the parameters of the first force field) and on the bases (with the parameters of the second force field). We have plotted potential energy curves for rotation around the  glycosidic bond $\chi$ for all four bases bound to the deoxyribose locked in the south conformation in a vacuum. It proved out that these curves lie between the corresponding curves for the CHARMM and AMBER force fields in the area of the $anti-$ conformation. 

For calculation of the electrostatic interactions, we used the PPPM (particle-particle particle-mesh Ewald) \cite{2021-computer-book-Hockney} method with an accuracy of six significant digits (the dimensionless LAMMPS parameter 10$^{-5}$). We chose to cut off the short-range part of the potential at the distance optimal for the calculation speed (8\AA{}). In the AMBER force field, we calculated the van der Waals forces with an accuracy of 10$^{-5}$~kcal/mol/\AA{} (i.e., we considered the force equal to zero at a distance larger than 8-20\AA{} depending on the type of atoms). In the CHARMM force field, we applied this approach only to ions and water (to their interactions with each other and with the DNA). The van der Waals interactions between the DNA atoms smoothly vanished in the range from 12 to 13\AA{}, regardless of the type of atoms.

We used the Langevin thermostat and the Nose-Hoover barostat (pressure of 1 atmosphere) with a damping parameter of one picosecond. We restrained X-H valence bonds and H-X-H valence angles (X is any atom) with 10$^{-5}$ kcal/mol tolerance by the SHAKE algorithm \cite{1977-SHAKE-algorithm}. The integration step was equal to 2~fs.

We set up the initial state of the system as follows. We generated the DNA molecule in the B-form by the code of the D.Case group \cite{D-Case-site}. We placed sodium and chloride ions at random positions around the DNA no closer than 3\AA{} to any other ion or DNA atom. Randomly oriented water molecules were placed in positions with oxygen atom not closer than 2.5\AA{} to any other atom in the system. Then we corrected the atom positions by energy minimization using the conjugate-gradient method (until the force vector norm became less than 10$^3$~kcal/mol/\AA). After that, we launched one  MD run at a temperature T=300K for 10 picoseconds in the NVT ensemble and then another run at the same temperature for 10 picoseconds in the NPT ensemble. One could start the MD relaxation of the system from this state, but the cloud of ions usually comes to equilibrium much later than other parts of such systems. In order to speed up this process, we performed an additional MD run with the fixed DNA molecule in the NPT ensemble for 2 nanoseconds. For the first nanosecond, the thermostat temperature for water molecules was 300K, and for ions - 500K. During the next nanosecond, we gradually cooled the ions (and the water if needed) down to the simulation temperature (300K, 283.15K, or 271.15K). The productive runs were carried out in the NPT ensemble.

In long-term all atom simulations, one observes too often and too long openings of the terminal base pairs \cite{2014-modeling-fraying,2015-modeling-against-fraying}. To avoid the fraying, we linked the bases in the terminal base pairs by a parabolic potential. More exactly, we replaced the energy of one of the hydrogen bonds of the pair (the central one: N1-H3 on A:T and N3-H1 on G:C) with a parabola. In the region where the energy reaches its minimum (2.2 - 2.3 \AA), the parabola approximates the sum of the van der Waals and Coulomb interactions between the nitrogen and the hydrogen atoms. For long oligomers, such restraints on the terminal base pairs should have little effect on the result of simulations.

We simulated the tetramer ATAT without the restraints on the terminal base pairs. It is known that at room temperature this oligomer is unstable and decomposes into two strands. At the experimental temperature of -2\textdegree C, when the oligomer is intact, in the force fields AMBER and 'AMBER-CHARMM's bases', the hydrogen bonds did not break during the simulation (55-219 ns in total from 4 replicas). In the CHARMM force field, the terminal hydrogen bonds broke after a few nanoseconds. In the case of the force  field 'CHARMM-AMBER's bases' (with stronger AMBER's base stacking), the breaks occurred less frequently and healed in most cases. However, in both the cases, we calculated the average values of the parameters Roll and Slide only on the central step of the oligomer ATAT.

In all the DNA force fields we used the same TIP4P/long water \cite{2004-TIP4P-Ew} and the same Joung-Cheatham ions \cite{2008-Young-Cheatham-ions}. Due to the specifics of the choice of partial charges in the CHARMM force field, the water model could have a significant impact on the calculation results. Therefore, we carried out several additional simulations with the inherent models for ions and water (file toppar\_water\_ions.str in CHARMM36). We simulated the oligomer GGTATACC with restrained hydrogen bonds on the terminal base pairs at two salt concentrations 0.15 and 4-5M. The replacement of the water and ions led to a decrease in the fraction of BII phosphate conformations, but only by 20\% from $\sim$60 to $\sim$40\%. The fraction of the north sugars increased from $\sim$3 to $\sim$15\% at the high salt concentration and up to 19\% at the low salt concentration. Both the changes were in the correct direction, but the observed values are still far from the experiment (17\% BII and $\sim$(20-50)\% C3'endo at 5-6M NaCl).

The computational cell was cubic, with periodic boundary conditions. The edge of the cell was about 15\AA{} greater than the axis of the extended oligomer in the B-form. We carried out the calculations for the oligomers with counterions, and in weak and strong salt solutions, with concentrations of approximately 0.1-0.15M and 4-5M. Table~\ref{table-simulations-summary} lists the total simulation time and the characteristics of the modeled systems depending on the salt concentration. In some cases, we collected statistics from several (usually 4) replicas. We stopped the calculation after the transition to another conformation (if any) was completed, and the lifetime in the new conformation was long enough to calculate the equilibrium average values and their dispersions for the parameters of the DNA helix.

\begin{table}
\caption{Parameters of the simulated systems: the number of water molecules and of pairs of additional salt ions for each concentration, and the total simulation time.
$^a$ Temperature -2\textdegree C; $^b$ Temperature 10\textdegree C
}
\label{table-simulations-summary}
\begin{tabular}{|c|c|c|c|c|}
\hline 
Sequence                               & NaCl, M & Water     & Ions   & Simulation time, ns\\
\hline
$\textrm{(GC)}_{10}$   & 0.15  & 19705   & 53     & 48-63\\
(polyGC)                         &           &              &         & \\\hline
$\textrm{A}_{20}$        & 0.15 & 19705 & 53      & 53-55\\
(polyA)                            & 4.1   & 17493 & 1412 & 54-59\\\hline                                      
ATAT$^a$                      & 0.13    & 1469  & 4         & 55-202\\
                                       & 4.5      & 1275 & 125        & 58-219\\\hline
GGTA-                           & 0.13   & 2864   & 7          & 140-224\\
TACC                             & 4.2       & 2562   & 196      & 149-234\\\hline
$\textrm{(AC)}_{10}$   & 0        & 19362 & 0         & 52-97\\
(polyAC)                         & 0.1     & 19308 & 34      & 51-152\\
                                         & 4.7     & 16895 & 1543 & 56-160\\\hline
$\textrm{G}_{20}$        & 0.15    & 19705 & 53     & 34-70\\
(polyG)                            & 1$^b$ & 19508 & 356     & 35-42\\
                                        & 4.1       & 17493 & 1412   & 25-101\\\hline
\end{tabular} 
\end{table}
\begin{acknowledgments}
This work was supported by the Program of Fundamental Research of the Russian Academy of Sciences (project FFZE-2022-0009). The calculations were carried out in the Joint Supercomputer Center of the Russian Academy of Sciences.
\end{acknowledgments}
\bibliography{Zubova-A-C-test}
\end{document}